\documentclass[preprint,aps]{revtex4}

\newcommand {\bea}{\begin{eqnarray}}
\newcommand {\eea}{\end{eqnarray}}
\newcommand {\be}{\begin{equation}}
\newcommand {\ee}{\end{equation}}

\newcommand {\Slash}{\!\!\!\!/\,}
\usepackage{graphicx}

\begin{document}


\title{Hard Loops, Soft Loops, and High Density Effective 
Field Theory}

\author{T.~Sch\"afer$^{1,2}$}

\affiliation{
$^1$Department of Physics, North Carolina State University,
Raleigh, NC 17695\\
$^2$Riken-BNL Research Center, Brookhaven National 
Laboratory, Upton, NY 11973}

\begin{abstract}
  We study several issues related to the use of effective 
field theories in QCD at large baryon density. We show that
the power counting is complicated by the appearance of 
two scales inside loop integrals. Hard dense loops involve
the large scale $\mu^2$ and lead to phenomena such as 
screening and damping at the scale $g\mu$. Soft loops
only involve small scales and lead to superfluidity 
and non-Fermi liquid behavior at exponentially small
scales. Four-fermion operators in the effective theory 
are suppressed by powers of $1/\mu$, but they get enhanced 
by hard loops. As a consequence their contribution to 
the pairing gap is only suppressed by powers of the 
coupling constant, and not powers of $1/\mu$. We determine 
the coefficients of four-fermion operators in the effective 
theory by matching quark-quark scattering amplitudes. Finally, 
we introduce a perturbative scheme for computing corrections 
to the gap parameter in the superfluid phase. 

\end{abstract}
\maketitle

\newpage

\section{Introduction}
\label{sec_intro}

  The study of baryonic matter in the regime of high baryon density 
has led to the discovery of new phases of strongly interacting matter,
such as color superconducting quark matter and color-flavor locked
matter \cite{Bailin:1984bm,Alford:1998zt,Rapp:1998zu,Alford:1999mk,Rajagopal:2000wf,Alford:2001dt,Schafer:2003vz,Rischke:2003mt}. 
These phases are not only relevant to the structure of compact 
astrophysical objects \cite{Reddy:2002ri,Alford:wf}, but they also 
provide a theoretical laboratory in which complicated QCD phenomena, 
such as chiral symmetry breaking and the formation of a mass gap, 
can be studied in a weakly coupled setting \cite{Schafer:1999ef}. 
In order to exploit these opportunities we would like to develop 
a systematic framework that will allow us to determine the exact 
nature of the phase diagram as a function of the density, temperature, 
the quark masses, and the lepton chemical potentials, and to compute 
the low energy properties of these phases. 

  If the density is large then the Fermi momentum is much bigger 
than the QCD scale, $p_F=\mu\gg\Lambda_{QCD}$, and it would seem that 
such a framework is provided by perturbative QCD. It is clear, 
however, that a naive expansion in powers of $\alpha_s$ is not
sufficient. First of all, it is well known that resummation 
is required in order to make the perturbative expansion in a 
many body system well defined \cite{GellMann:1957}. It is also 
well known that additional problems arise in systems with unscreened 
transverse gauge boson interactions \cite{Holstein:1973,Reizer:1989}. 
In a degenerate Fermi system the effect of the BCS or other pairing 
instabilities have to be taken into account. And finally, in systems
with broken global symmetries, the low energy properties of the 
system are governed by collective modes that carry the 
quantum numbers of the broken generators. 

  In order to address these problems it is natural to exploit 
the separation of scales provided by $\mu\gg g\mu \gg\Lambda_{QCD}$ 
in the normal phase, or $\mu\gg g\mu \gg\Delta\gg\Lambda_{QCD}$ in 
the superfluid phase. An effective field theory approach to 
phenomena near the Fermi surface was suggested by Hong
\cite{Hong:2000tn,Hong:2000ru}. This approach was applied
to the gap equation \cite{Beane:2000ji}, Goldstone boson 
masses \cite{Beane:2000ms,Schafer:2001za}, gluon dispersion 
relations \cite{Casalbuoni:2001ha}, and a number of other problems.
For a review and further references, see \cite{Nardulli:2002ma}.
Even though a number of interesting results have been obtained
there are a number of important conceptual issues that are not 
very well understood. These issues concern power counting,
renormalization and matching. In this work we would like to 
study some of these issues in more detail. This paper is 
organized as follows. In section \ref{sec_hdl} we give a 
review of the standard hard thermal loop approximation applied 
to dense matter. In Section \ref{sec_hdet} we introduce the 
high density and effective theory (HDET) and in Section 
\ref{sec_hloop} we show how hard dense loops (HDLs) arise 
in the effective theory. In Section \ref{sec_sloop} we show 
that the effective theory contains a new class of diagrams which 
we shall call soft dense loops. In Sections \ref{sec_gap} and 
\ref{sec_pgap} we study the gap equation at leading and 
next-to-leading order and in Section \ref{sec_match} we 
study matching conditions for four-fermion operators. 

\section{Hard Dense Loops}
\label{sec_hdl}

 In this section we review the calculation of hard dense loop 
contributions to QCD Greens functions 
\cite{Braaten:1989mz,Blaizot:1993bb,Manuel:1995td}. 
The hard dense loop limit corresponds to soft external 
momenta, $|\vec{p}|\ll\mu$. In this limit the main 
medium contribution comes from hard loop momenta, 
$|\vec{k}|\sim \mu$. Our main purpose in this section is 
to present a simple rederivation of the main results that will 
allow us to compare the hard dense loop approximation with the 
high density effective theory which is discussed in section 
\ref{sec_hdet}. 

 Hard dense loops can be calculated most simply by writing the 
free fermion propagator at $\mu\neq 0$ in the form 
\be
\label{hdl_s}
S(k) = \frac{1}{2}\left\{\frac{\hat{K}\Slash }{k_0-\epsilon^+_k}
  + \frac{\hat{\bar{K}}\Slash }{k_0-\epsilon^-_k}
   \right\},
\ee
where $\epsilon_k^\pm = \pm(|\vec{k}|\mp\mu)$ and $\hat{K}
=(1,\hat{k})$, $\hat{\bar{K}}=(1,-\hat{k})$ with $\hat{k}
=\vec{k}/|\vec{k}|$. The two poles of the propagator in 
equ.~(\ref{hdl_s}) correspond to particle and anti-particle 
states. Note that $\Lambda_k^+=\hat{K}\Slash\gamma_0/2$ and
$\Lambda_k^-=\hat{\bar{K}}\Slash\gamma_0/2$ are projection
operators on positive and negative energy solutions of the 
free Dirac equation. Also observe that $(\hat{K}\Slash )^2
=(\hat{\bar{K}}\Slash)^2=0$ which is very useful in 
evaluating Dirac traces in the hard dense loop 
approximation. In our convention the energy is measured
relative to the Fermi energy $\mu$. In order to derive 
HDL Greens functions it is useful to define the energy 
of a fermion as $\omega=k_0+\mu$. The HDL limit corresponds 
to $\omega,|\vec{k}|\ll\mu$, whereas quasi-particles in 
the vicinity of the Fermi surface satisfy $\omega,
|\vec{k}|\sim\mu$. 

  As an example, let us consider the calculation of the 
gluon polarization function. The one-loop contribution
is given by 
\be 
\Pi_{\mu\nu}^{ab}(p)  = 
 N_fg^2\frac{\delta^{ab}}{2}
 \int\frac{d^4k}{(2\pi)^4} 
  {\rm Tr}\left( \gamma_\mu S(k)\gamma_\nu S(k-p)\right).
\ee
If the propagator is written in terms of particle and 
anti-particle contributions as in equ.~(\ref{S_hdl}) we find 
three contributions to the polarization function, the particle-hole, 
particle-anti-particle, and anti-particle-anti-hole terms. 
The particle-hole contribution shown in Fig.\ref{fig_hdl1}a is 
\be
\label{pi_ph}
\left.\Pi_{\mu\nu}^{ab}(p)\right|_{ph} = 
 2m^2 \delta^{ab}\int \frac{d\Omega}{4\pi}
 \hat{K}_\mu\hat{K}_\nu
 \left\{ 1-\frac{p_0}{p\cdot\hat{K}} \right\}
\ee
where $m^2=N_fg^2\mu^2/(4\pi^2)$. This contribution is 
dominated by loop momenta $|\vec{k}|\sim \mu$ very close
to the Fermi surface. However, the particle-hole term is 
not transverse \cite{Hong:2000ru,Rischke:2000qz}. The 
particle-anti-particle term is 
\be 
\label{pi_pa}
\left.\Pi_{\mu\nu}^{ab}(p)\right|_{pa} = 
2m^2 \delta^{ab}\int \frac{d\Omega}{4\pi} 
 \left\{ \delta_{\mu 0}\delta_{\nu 0} -
\hat{K}_\mu\hat{K}_\nu \right\}.
\ee
Note that this term receives contributions from particles
with momenta in the range $0\leq |\vec{k}|\leq\mu$, and not 
only particles in the vicinity of the Fermi surface. Also note 
that in the hard dense loop limit this term is just a constant, 
with no momentum dependence. The sum of the two contributions 
equ.~(\ref{pi_ph}) and (\ref{pi_pa}) is given by
\be
\label{pi_hdet}
\Pi_{\mu\nu}^{ab}(p) = 
 2m^2 \delta^{ab}\int \frac{d\Omega}{4\pi} 
 \left\{ \delta_{\mu 0}\delta_{\nu 0}-
\frac{p_0\hat{K}_\mu\hat{K}_\nu}{p\cdot\hat{K}} \right\}.
\ee
The polarization tensor equ.~{\ref{pi_hdet}) is transverse
and agrees with the well known hard dense loop result. 

 The fermion self energy in the hard dense loop approximation
is given by \cite{Blaizot:1993bb}
\be 
\label{Sigma_f}
\Sigma(p) = m_f^2 \int \frac{d\Omega}{4\pi} 
 \frac{\hat{K}\Slash}{p\cdot\hat{K}},
\ee
with $m_f^2=C_Fg^2\mu^2/(8\pi^2)$ and $C_F=(N_c^2-1)/(2N_c)$. This 
result arises solely from the particle term in the fermion propagator 
equ.~(\ref{S_hdl}). The loop integral receives contributions from 
all momenta $0\leq |\vec{k}|\leq\mu$. It is important to emphasize 
that the HDL approximation corresponds to external momenta that 
are soft, $|\vec{p}|\ll\mu$, whereas low energy fermionic 
excitations have momenta $|\vec{p}|\sim \mu$. The use of the 
HDL fermion self energy is not reliable for low energy modes.
It is nevertheless instructive to study the behavior of the
HDL self energy near the Fermi surface. Using equ.~(\ref{Sigma_f})
we find
\be 
\Sigma(p) = \Sigma_S(p)-\Sigma_V(p)\vec{\gamma}\cdot\hat{p}
\ee
with 
\bea 
\Sigma_S(p) &=& \frac{m_f^2}{|\vec{p}|}
  Q_0\left(\frac{\omega}{|\vec{p}|}\right), 
 \hspace{1.5cm} Q_0\left(\frac{\omega}{|\vec{p}|}\right) = 
  \frac{1}{2}\log\left(\frac{\omega+|\vec{p}|}
   {\omega-|\vec{p}|}\right),\\
\Sigma_S(p) &=& -\frac{m_f^2}{|\vec{p}|}
 \left(1-\frac{\omega}{|\vec{p}|}
  Q_0\left(\frac{\omega}{|\vec{p}|}\right)\right).
\eea
We observe that both $\Sigma_S$ and $\Sigma_V$ are logarithmically
divergent near the Fermi surface, but the quasi-particle dispersion
relation and the wave function renormalization factor are well
behaved. Indeed, for the particle mode
\be
S(p)=(S_0^{-1}(p)-\Sigma(p))^{-1} \simeq
  \frac{Z_+(p)}{p_0-\omega^+_p} \frac{\hat{P}\Slash}{2}
  +\ldots, 
\ee
with    
\be
\omega_+(p)\simeq|\vec{p}|+\frac{m_f^2}{|\vec{p}|},
\hspace{0.5cm}
Z_+(p)\simeq 1+\frac{m_f^2}{2|\vec{p}|^2}\left(1-\log\left(
  \frac{2|\vec{p}|^2}{m_f^2}\right)\right).
\ee

 At one-loop order the quark-gluon vertex receives contributions
from the abelian diagram Fig.~\ref{fig_hdl2}a and the non-abelian
diagram Fig.~\ref{fig_hdl2}b. In the HDL limit these two diagrams 
are identical up to a color factor. We find
\be 
\label{vert_hdl}
\Gamma_\mu(p_1,p_2) = -m_f^2
\int\frac{d\Omega}{4\pi} \frac{\hat{K}_\mu\hat{K}\Slash}
{(p_1\cdot\hat{K})(p_2\cdot\hat{K})},
\ee
where $p_1,p_2$ are the momenta of the two fermions and we have 
factored out the color structure $\lambda^a/2$. The vertex
correction is similar to the fermion self energy in the sense
that the loop integral receives contribution from momenta 
$0\leq|\vec{k}|\leq\mu$. Also, we can study the effect of 
the HDL effective vertex for quasi-particles in the vicinity
of the Fermi surface. Since both the incoming and outgoing 
Fermions are hard the interesting regime corresponds to 
soft gluon momenta. In this limit the matrix element of the
free vertex between projectors on quasi-particle states with 
momentum $p$ is $\langle\gamma_\mu \rangle = \hat{P}_\mu$. 
For the HDL vertex correction we find 
\be 
\langle \Gamma_\mu (p,p')\rangle = 
 -\frac{m_f^2}{\mu}\int\frac{d\Omega}{4\pi}
 \frac{\hat{K}_\mu}{p'\cdot\hat{K}},
\ee
which has a logarithmic divergence near the Fermi surface. 

 The gluonic three-point function in the HDL limit only 
receives contributions from the fermion loop diagram shown 
in Fig.~\ref{fig_hdl3}a. We find
\be 
\label{gam_hdl}
\Gamma^{abc}_{\mu\nu\alpha}(p,q,r) = igf^{abc} 2m^2 
\int \frac{d\Omega}{4\pi} \hat{K}_\mu\hat{K}_\alpha
 \hat{K}_\beta \left\{ 
  \frac{q_0}{(q\cdot\hat{K})(p\cdot\hat{K})}
 -\frac{r_0}{(r\cdot\hat{K})(p\cdot\hat{K})}
 \right\}.
\ee
The three-point function, as well as higher $n$-point functions,
is dominated by momenta near the Fermi surface. Also, as 
opposed to the case of the two-point function, anti-particles
only make sub-leading contributions. Higher $n$-point functions
can be computed directly using the methods discussed here. 
Alternatively, as is the case for $T\neq 0$, $n\geq3$ point 
functions can be reconstructed using Ward identities or 
non-local effective actions. To summarize this section, we
showed that gluon $n$-point functions in the HDL limit are
dominated by modes near the Fermi surface. The only exception
is the two-point function which requires a contact term 
determined by modes with momenta between zero and the Fermi
momentum. Fermion HDLs, on the other hand, are always 
determined by momenta in the range $0\leq|\vec{k}|\leq\mu$.
We emphasized, however, that the HDL limit is not relevant
for low energy fermionic excitations.

\section{High Density Effective Theory (HDET)}
\label{sec_hdet}

 In this work we wish to compare the hard dense loop results 
with the Greens functions obtained from an effective theory which 
describes low energy excitations in the vicinity of the Fermi 
surface. The leading order terms in the effective theory are
\cite{Hong:2000tn,Hong:2000ru,Beane:2000ms}
\be
\label{fs_eff}
{\cal L} =\sum_{v}
 \psi_{v}^\dagger (iv\cdot D) \psi_{v} 
 -\frac{1}{4}G^a_{\mu\nu} G^a_{\mu\nu}+ \ldots ,
\ee
where $v_\mu=(1,\vec{v})$. In the vicinity of the Fermi surface 
the relevant degrees of freedom are particle and hole excitations 
which move with the Fermi velocity $v$. We shall describe these 
excitations in terms of a field $\psi_{v}(x)$. This field 
describes particles and holes with momenta $p=\mu\vec{v}+l$, where 
$l\ll\mu$. We will write $l=l_0+l_{\|}+l_\perp$ with $\vec{l}_{\|}
=\vec{v}(\vec{l}\cdot \vec{v})$ and $\vec{l}_\perp = 
\vec{l}-\vec{l}_{\|}$. In order to take into account the entire 
Fermi surface we have to cover the Fermi surface with patches 
labeled by the local Fermi velocity, see Fig.~\ref{fig_fs}. The 
number of such patches is $n_v\sim (\mu^2/\Lambda_\perp^2)$ where 
$\Lambda_\perp \ll\mu$ is the cutoff on the transverse momenta 
$l_\perp$. 

 Higher order terms are suppressed by powers of $1/\mu$. As 
usual we have to consider all possible terms allowed by the 
symmetries of the underlying theory. At $O(1/\mu)$ we have
\be
{\cal L} = \sum_{v}\left\{
  -\frac{1}{2\mu} \psi_{v}^\dagger D_\perp^2 \psi_{v}
  - ag  \psi_{v}^\dagger \frac{\sigma^{\mu\nu} G_{\mu\nu}^\perp}
   {4\mu}\psi_v \right\}. 
\ee
The coefficient of the first term is fixed by the dispersion 
relation of a fermion near the Fermi surface, $l_0=l_{\|}+
l_\perp^2/(2\mu)+\ldots$. The coefficient of the second term is 
most easily determined by writing the quark field in the microscopic 
theory as $\psi=\psi_++\psi_-$ and then integrating out $\psi_-$ at 
tree level. We find $a=1+O(g^2)$, where the $O(g^2)$ terms 
arise from higher order perturbative corrections. At higher 
order in $1/\mu$ there is an infinite tower of operators 
of the form $\mu^{-n}\psi^\dagger_v D_\perp^{2n_1}(\bar{v}
\cdot D)^{n_2} \psi_v$ with $\bar{v}=(1,-\vec{v})$ and $n=
2n_1+n_2-1$. 

At $O(1/\mu^2)$ the effective theory contains four-fermion 
operators 
\be 
{\cal L}=  \frac{1}{\mu^2} \sum_{v_i} \sum_{\Gamma,\Gamma'} 
  c^{\Gamma\Gamma'}(\vec{v}_1\cdot\vec{v}_2,\vec{v}_1\cdot\vec{v}_3,
   \vec{v}_2\cdot\vec{v}_3) 
   \Big(\psi_{v_1} \Gamma \psi_{v_2}\Big)
   \Big(\psi^\dagger_{v_3}\Gamma'\psi^\dagger_{v_4}\Big)
   \delta_{v_1+v_2-v_3-v_4}.
\ee
The restriction $v_1+v_2=v_3+v_4$ allows two types of four-fermion 
operators. The first possibility is that both the incoming and 
outgoing fermion momenta are back-to-back. This corresponds to
the BCS interaction
\be
\label{c_bcs}
{\cal L}=  \frac{1}{\mu^2}\sum_{v,v'}\sum_{\Gamma,\Gamma'}
  V_l^{\Gamma\Gamma'} R_l^{\Gamma\Gamma'}(\vec{v}\cdot\vec{v}') 
    \Big(\psi_{v} \Gamma \psi_{-v}\Big)
   \Big(\psi^\dagger_{v'}\Gamma'\psi^\dagger_{-v'}\Big),
\ee
where $\vec{v}\cdot\vec{v}'=\cos\theta$ is the scattering 
angle and $R_l^{\Gamma\Gamma'}(\vec{v}\cdot\vec{v}')$ is a 
set of orthogonal polynomials that we will specify below.
The second possibility is that the final momenta are equal 
to the initial momenta up to a rotation around the axis 
defined by the sum of the incoming momenta. The relevant 
four-fermion operator is 
\be
\label{c_flp}
{\cal L}=  \frac{1}{\mu^2}\sum_{v,v',\phi}\sum_{\Gamma,\Gamma'}
  F_l^{\Gamma\Gamma'}(\phi) R_l^{\Gamma\Gamma'}(\vec{v}\cdot\vec{v}') 
    \Big(\psi_{v} \Gamma \psi_{v'}\Big)
   \Big(\psi^\dagger_{\tilde{v}}\Gamma'\psi^\dagger_{\tilde{v}'}
  \Big),
\ee
where $\tilde{v},\tilde{v}'$ are the vectors obtained from 
$v,v'$ by a rotation around $v_{tot}=v+v'$ by the angle $\phi$.
In a system with short range interactions only the quantities 
$F_l(0)$ are known as Fermi liquid parameters. The matrices 
$\Gamma,\Gamma'$ describe the spin, color and flavor structure 
of the interaction. In the following we will focus on the 
spin structure. We can decompose 
\be
\Big(\Gamma\Big)\Big(\Gamma'\Big)=\left\{
 \Big(H_\pm^T \sigma_2 H_\pm\Big)
  \Big(H_\pm^T \sigma_2 H_\pm\Big),
 \Big(H_\pm^T \sigma_2\vec{\sigma} H_\pm\Big)
  \Big(H_\pm^T \sigma_2\vec{\sigma} H_\pm\Big)\right\},
\ee
where $H_\pm=(1\pm\vec{v}\cdot\vec{\sigma})/2$ are helicity
projectors. In the spin zero sector there are two possible
helicity channels, $(++)\to(++)$ and $(++)\to (--)$ together 
with their parity partners $(+\leftrightarrow -)$. In the limit
$m\to 0$ perturbative interactions only contribute to the 
helicity non-flip amplitude
\be
\label{v_bcs_0}
{\cal L}=  \frac{1}{\mu^2}\sum_{v,v'}
  V_l^{++} P_l(\vec{v}\cdot\vec{v}') 
    \Big(\psi_{v} \sigma_2 H_+\psi_{-v}\Big)
   \Big(\psi^\dagger_{v'}\sigma_2 H_+\psi^\dagger_{-v'}\Big)
 + (+\leftrightarrow -),
\ee
where $P_l(\cos\theta)$ are Legendre polynomials. At $O(m^2/
\mu^2)$ quark mass terms induce a non-zero helicity flip amplitude.
The corresponding four-fermion operator was determined in 
\cite{Schafer:2001za}. In $N_f=2$ QCD instantons generate 
a helicity-flip amplitude which is suppressed by extra 
powers of $(\Lambda_{QCD}/\mu)$ \cite{Schafer:1999na}. In 
QCD with $N_f=3$ flavors instantons produce a helicity changing 
four-fermion interaction which is suppressed by both 
$(\Lambda_{QCD}/\mu)$ and $(m/\mu)$ \cite{Schafer:2002ty}. 
Even though helicity changing amplitudes are suppressed, 
they have important physical effects. For example, helicity
flip amplitudes determine the masses of Goldstone bosons in 
the CFL and 2SC phase. 

 In the spin one sector there is only one helicity channel 
$(+-)\to(+-)$. The corresponding BCS interaction is 
\be
\label{v_bcs_1}
{\cal L}=  \frac{1}{\mu^2}\sum_{v,v'}
  V_l^{+-} d^{(l)}_{11}(\vec{v}\cdot\vec{v}') 
    \Big(\psi_{v} \sigma_2H_-\vec{\sigma} H_+\psi_{-v}\Big)
   \Big(\psi^\dagger_{v'}\sigma_2H_-\vec{\sigma} H_+
    \psi^\dagger_{-v'}\Big)
 + (+\leftrightarrow -),
\ee
where $d^{(l)}_{11}(\cos\theta)$ is the reduced Wigner D-function.
For details on the helicity amplitude formalism we refer the 
reader to \cite{Jacob:at}. Note that in the total helicity 
zero channel the reduced rotation matrix reduces to a 
Legendre polynomial, $d^{(l)}_{00}(\cos\theta)=P_l(\cos\theta)$, 
in agreement with equ.~(\ref{v_bcs_0}). We will discuss the 
matching conditions for the the spin zero and spin one BCS 
helicity amplitudes $V_l^{++}$ and $V_l^{+-}$ 
in Sect.~\ref{sec_match}.

\section{Hard Loops}
\label{sec_hloop}

 There are two types of loop diagrams that arise in 
high density effective field theory, hard dense loops 
and soft dense loops. As an example of a hard dense loop
we study the gluon two point function. At leading order in $g$ 
and $1/\mu$ we have 
\be 
\Pi^{ab}_{\mu\nu}(p) = 2g^2N_f\frac{\delta^{ab}}{2}
  \sum_{\vec{v}} v_\mu v_\nu 
 \int \frac{d^4k}{(2\pi)^4} 
 \frac{1}{(k_0-l_k)(k_0+p_0-l_{k+p})},
\ee
where $l_k=\vec{v}\cdot \vec{k}$. We note that taking the momentum
of the external gluon to zero automatically selects forward scattering.
We also observe that the gluon can interact with fermions of any 
Fermi velocity so that the polarization function involves a sum over 
all patches. After performing the $k_0$ integration we get 
\be 
\Pi^{ab}_{\mu\nu}(p) = 
 2g^2N_f\frac{\delta^{ab}}{2}
  \sum_{\vec{v}} v_\mu v_\nu 
 \int \frac{d^2l_\perp}{(2\pi)^2}
 \int \frac{dl_k}{2\pi} \frac{l_p}{p_0-l_p}
  \frac{\partial n_k}{\partial l_k},
\ee
where $n_k$ is the Fermi distribution function. We note 
that the $l_k$ integration is automatically restricted to
small momenta. The integral over the transverse momenta
$l_\perp$, on the other hand, diverges quadratically with
the cutoff $\Lambda_\perp$. We observe, however, that the 
sum over patches and the integral over $l_\perp$ can 
be combined into an integral over the entire Fermi
surface
\be
\label{hard_int}
 \frac{1}{2\pi}\sum_{\vec{v}}\int\frac{d^2l_\perp}{(2\pi)^2}
 =\frac{\mu^2}{2\pi^2}\int \frac{d\Omega}{4\pi}.
\ee
This identification is consistent with our observation that 
the number of patches is $n\sim \mu^2/(2\pi^2\Lambda_\perp^2)$.
As a consequence of equ.~(\ref{hard_int}) the upper limit of 
the $l_\perp$ integral is effectively $\mu$, and not 
$\Lambda_\perp\ll\mu$. We will refer to loop integrals 
of this kind as hard loops. We find
\be
\label{pi_hdet2}
\Pi^{ab}_{\mu\nu}(p) = 2m^2\delta^{ab} \int\frac{d\Omega}{4\pi}
  v_\mu v_\nu \left\{ 1-\frac{p_0}{p_0-l_p}
 \right\}.
\ee
This result agrees with equ.~(\ref{pi_ph}). We know however, 
that equ.~(\ref{pi_hdet2}) is not transverse and that the 
contribution from anti-particles, equ.~(\ref{pi_pa}) is 
missing. In the effective theory, this contribution has 
to represented by a local counterterm. The required 
counterterm is \cite{Hong:2000tn}
\be
{\cal L}= \frac{1}{2} m^2 \int\frac{d\Omega}{4\pi}
 (\vec{A}_\perp)^2.
\ee
The appearance of this term is related to the fact that terms 
in the lagrangian of the form $\mu^{-n}\psi^\dagger_v D_\perp^{2n_1}
(\bar{v}\cdot D)^{n_2} \psi_v$ give non-vanishing tadpole contributions
proportional to $g^{2n_1+n_2}\mu^{4-2n_1-n_2} A_\perp^{2n_1}
(\bar{v}\cdot A)^{n2}$ which have to be absorbed into counterterms. 

Putting everything together we find
\be
\label{pi_hdet3}
\Pi_{\mu\nu}(p) = 2m^2 \int\frac{d\Omega}{4\pi}
   \left\{ \delta_{\mu 0}\delta_{\nu 0} -
   \frac{v_\mu v_\nu p_0}{p_0-l_p}
 \right\}
\ee
which agrees with the complete HDL result equ.~(\ref{pi_hdet}).
The gluonic three-point function can be computed in the 
same fashion. Using the rule equ.~(\ref{hard_int}) we find
\be 
\label{gam_hdet}
\Gamma^{abc}_{\mu\nu\alpha}(p,q,r) = igf^{abc} 2m^2 
\int \frac{d\Omega}{4\pi} v_\mu v_\alpha v_\beta \left\{ 
  \frac{q_0}{(q\cdot v)(p\cdot v)}
 -\frac{r_0}{(r\cdot v)(p\cdot v)}
 \right\},
\ee
which agrees with the HDL result equ.~(\ref{gam_hdl}). We 
note that in the case of the three point function, as well
as higher $n$-point functions, there is no leading order 
contribution from anti-particle states, and this is reflected
by the absence of counterterms at leading order in the 
effective theory.

\section{Soft Loops}
\label{sec_sloop}

 As an example of a soft loop contribution in the high 
density effective theory we study the fermion self energy.
At leading order, we have
\be
\label{sigma_0}
\Sigma(p) = g^2 C_F\int\frac{d^4k}{(2\pi)^4}
  \frac{1}{p_0+k_0-l_{p+k}} v_\mu v_\nu D_{\mu\nu}(k),
\ee
where $D_{\mu\nu}(k)$ is the gluon propagator. Soft contributions
to the quark self energy are dominated by nearly forward scattering. 
Note that this loop integral does not involve a sum over patches. 
Hard contributions to the fermion self energy are governed 
by the four-fermion operators equ.~(\ref{c_flp}). We saw in 
the previous section that hard loops cause non-perturbative 
effects in gluon $n$-point functions at the scale $g\mu$. At 
energies below this scale we have to replace the free gluon 
term in the effective lagrangian by the generating functional 
for hard dense loops \cite{Braaten:1991gm,Braaten:1992jj}
\be 
\label{S_hdl}
{\cal L} = -m^2\int\frac{d\Omega}{4\pi}
 {\rm Tr}\,G_{\mu \alpha} 
  \frac{\hat{P}^\alpha \hat{P}^\beta}{(\hat{P}\cdot D)^2} 
G^\mu_{\,\beta},
\ee
where the angular integral corresponds to an average over 
the direction of $\hat{P}_\alpha =(1,\hat{p})$. Note 
that the effective gluonic action is non-local. The effective 
fermion action in the high density effective theory for momenta
below $g\mu$ remains local. The hard dense loop effective 
action equ.~(\ref{S_hdl}) leads to the gluon propagator 
\be
D_{\mu\nu}(k) = \frac{P_{\mu\nu}^T}{k^2-\Pi_M} 
 + \frac{P_{\mu\nu}^L}{k^2-\Pi_E}
\ee
where $\Pi_M$ and $\Pi_E$ are the transverse and longitudinal 
self energies in the HDL limit. The projection operators 
$P_{\mu\nu}^{T,L}$ are defined by 
\bea
\label{proj}
 P_{ij}^T &=& \delta_{ij}-\hat{k}_i\hat{k}_j , \hspace{1cm}
 P_{00}^T = P_{0i}^T = 0,    \\
 P_{\mu\nu}^L &=& -g_{\mu\nu}+\frac{k_\mu k_\nu}{k^2}
   - P_{\mu\nu}^T .
\eea
In the regime $|k_0|<|\vec{k}|<g\mu$ the self energies can be 
approximated by $\Pi_E = 2m^2$ and  $\Pi_M = i\frac{\pi}{2}m^2
k_0/|\vec{k}|$. We note that in this regime the transverse
self energy is much smaller than the longitudinal one, $\Pi_M<
\Pi_E$. As a consequence the dominant part of the fermion self
energy arises from transverse gluons. We have 
\be 
\label{sigma_1}
\Sigma(p) = g^2C_F \int \frac{dl_0}{2\pi}\int \frac{l^2dl}{(2\pi)^2}
\int_{-1}^1 dx \frac{1-x^2}{p_0+l_0-l_p-lx}
 \frac{1}{l_0^2-l^2+i\frac{\pi}{2}m^2 \frac{l_0}{l}},
\ee
where $l_p=\vec{v}\cdot\vec{p}-\mu$ and $l_k=\vec{v}\cdot\vec{k}
\equiv  lx$. To leading logarithmic accuracy we can ignore 
the difference between transverse and longitudinal cutoffs 
and set $\Lambda_\perp=\Lambda_{\|}=\Lambda$. We can also 
set $l_p\to 0$ and $(1-x^2)\to 1$. We compute the integral 
by analytic continuation to euclidean space. Performing 
the integral over $x$ we have 
\be 
\Sigma(p) = 2C_Fg^2 \int \frac{dl_4}{2\pi}\int \frac{ldl}{(2\pi)^2}
 \arctan\left(\frac{l}{p_4+l_4}\right) 
 \frac{1}{l_4^2+l^2+\frac{\pi}{2}m^2 \frac{l_4}{l}}.
\ee
The leading logarithmic term in the energy $p_4$ can be extracted 
from 
\be
\label{sigma_2}
\frac{d}{dp_4}\Sigma(p_4) =  2g^2C_F 
 \int \frac{dl_4}{2\pi}\int \frac{ldl}{(2\pi)^2}
 \left\{ \pi \delta (l_4+p_4) -\frac{l}{(l_4+p_4)^2+l^2}
 \right\}
\frac{1}{l_4^2+l^2+\frac{\pi}{2}m^2 \frac{l_4}{l}}.
\ee
Only the first term in the curly brackets gives a 
logarithmic contribution in the limit $p_4\to 0$. We get 
\cite{Brown:1999yd,Brown:2000eh,Boyanovsky:2000bc,Manuel:2000nh,Vanderheyden:1996bw}
\be
\label{sigma_3}
\Sigma(p_4) \simeq \frac{g^2C_Fp_4}{4\pi^2} 
 \int dl \frac{l}{l^2+\frac{\pi}{2}m^2 \frac{p_4}{l}}
 \simeq  \frac{g^2C_F}{12\pi^2}p_4 \log\left( 
  \frac{\Lambda}{p_4}\right) ,
\ee
which is accurate up to terms of $O(g^2p_4)$. Without calculating 
the $O(g^2p_4)$ term we cannot fix the scale inside the logarithm
in equ.~(\ref{sigma_3}). 

Also note that the soft fermion self energy is of natural size.
The soft gluon propagator scales as $1/\Lambda_{soft}^2$, the 
soft fermion propagator scales as $1/\Lambda_{soft}$, and the 
loop integral gives $\Lambda_{soft}^4$. As a result, the expected
scaling is $\Sigma\sim \Lambda_{soft}$ which agrees with 
equ.~(\ref{sigma_3}) since $p_4$ is a soft momentum. For 
comparison, the loop integral in the hard loop diagram discussed 
in section \ref{sec_hloop} scales as $\mu^2\Lambda^2_{soft}$. 

 Ward identities relate the soft fermion self energy to the 
soft quark-gluon vertex. The leading contribution to the 
abelian vertex function is given by
\be
\label{vertex_1}
\Gamma^{(ab)}_\alpha (p_1,p_2) = g^3C^{(ab)}\int\frac{d^4k}{(2\pi)^4}
v_\mu v_\alpha v_\nu D_{\mu\nu}(k)
 \frac{1}{(p_1-k)_0-l_{p_1-k}}
 \frac{1}{(p_2-k)_0-l_{p_2-k}},
\ee
with $C^{(ab)}=C_F-C_A/2$ and $C_A=N_c$. As in the case of the 
quark self energy the dominant contribution arises from the transverse 
part of the gluon propagator. Continuing to euclidean space we find
\bea
\label{vertex_2}
\Gamma^{(ab)}_\alpha (p_1,p_2) &=& g^3C^{(ab)}v_\alpha 
\int \frac{dl_4}{2\pi}\int \frac{l^2dl}{(2\pi)^2}
\int_{-1}^1 dx 
 \frac{1-x^2}{l_4^2+l^2+\frac{\pi}{2}m^2 \frac{l_4}{l}}
 \nonumber \\
& & \hspace{4cm}\cdot 
 \frac{1}{i(p_1-l)_4-l_{p_1}-lx}
 \frac{1}{i(p_2-l)_4-l_{p_2}-lx}.
\eea
One can check that the term proportional to $x^2$ can be 
neglected. Performing the integral over $x$ we get
\bea
\label{vertex_3}
\Gamma^{(ab)}_\alpha (p_1,p_2) &=&  g^3C^{(ab)}v_\alpha 
\int \frac{dl_4}{2\pi}\int \frac{ldl}{(2\pi)^2}
\frac{1}{l_4^2+l^2+\frac{\pi}{2}m^2 \frac{l_4}{l}}
\frac{1}{i(p_2-p_1)_4-(l_{p_2}-l_{p_1})} \nonumber \\
& & \hspace{1cm}
\left\{ \log\left(
 \frac{i(p_1-l)_4-(l_{p_1}+l)}{i(p_1-l)_4-(l_{p_1}-l)} \right)
-\log\left(
 \frac{i(p_2-l)_4-(l_{p_2}+l)}{i(p_2-l)_4-(l_{p_2}-l)} \right)
  \right\}.
\eea
Brown et al.~observed that there are two distinct kinematic 
regimes for the vertex function, depending on whether $(p_1-p_2)_4$
is bigger or smaller than $(l_{p_1}-l_{p_2})$ \cite{Brown:2000eh}. 
In the limit $(l_{p_1}-l_{p_2})\ll(p_1-p_2)_4$ we get 
\be 
\label{vertex_4}
\Gamma^{(ab)}_\alpha (p_1,p_2) = g^3C^{(ab)} v_\alpha 
\int \frac{dl_4}{2\pi}\int \frac{ldl}{(2\pi)^2}
\left\{ \pi \delta (l_4+p_4) -\frac{l}{(l_4+p_4)^2+l^2}
 \right\}
\frac{1}{l_4^2+l^2+\frac{\pi}{2}m^2 \frac{l_4}{l}},
\ee
where $(p_1)_4 \simeq (p_2)_4\simeq p_4$. This integral is clearly 
identical to equ.~(\ref{sigma_2}) for the derivative of the fermion 
self energy. We get
\be 
\label{vertex_5} 
\lim_{(p_1)_4\to (p_2)_4} \lim_{l_{p_1}\to l_{p_2}}
\Gamma^{(ab)}_\alpha (p_1,p_2) = 
\frac{g^3C^{(ab)} v_\alpha}{12\pi^2}
\log\left(\frac{\Lambda}{p_4}\right).
\ee
In the opposite limit $(l_{p_1}-l_{p_2})\gg(p_1-p_2)_4$ the 
delta function is not present and the vertex function does 
not have a logarithmic divergence. 

 We saw that in the hard dense loop approximation the abelian 
and non-abelian vertex corrections are identical up to a color
factor. The sum of the color factors of the two diagrams is 
$C^{(ab)}+C^{(nab)}=(C_F-C_A/2)+C_A/2=C_F$ which shows that 
the quark-gluon vertex in the HDL approximation has the same
color structure as the quark self energy. The same result is 
also obtained for the leading IR divergence in the soft dense 
loop limit. The only difference is that following the arguments 
given above we have to take into account both the HDL gluon 
self energy and the HDL gluon three-point function when evaluating 
the non-abelian vertex function, see Fig.~\ref{fig_sdl}c. 
Indeed, the free gluon three-point function does not 
contribute to the IR divergence of the quark-gluon vertex 
in the SDL limit. We have 
\bea
\label{vertex_6}
\Gamma^{(nab)}_\alpha (p_1,p_2) &=& g^3C^{(nab)}\int\frac{d^4k}{(2\pi)^4}
v_\mu v_\nu D_{\mu\mu'}(k)
 \Gamma_{\alpha\mu'\nu'}(p_2-p_1,k,p_1-p_2-k) \nonumber \\
 & & \hspace{2cm} \cdot D_{\nu'\nu}(p_1-p_2-k) 
  \frac{1}{(p_1)_0-k_0-l_{p_1-k}}.
\eea
Again the interesting limit is $(l_{p_1}-l_{p_2})\ll(p_1-p_2)_4$.
The calculation is greatly simplified by the observation that 
in this limit equ.~(\ref{gam_hdet}) gives 
\bea
\label{gam_p4}
\Gamma_{4\mu\nu}(p_2-p_1,k,p_1-p_2-k) &\simeq &
 \frac{1}{(p_2-p_1)_4}
 \left\{ \Pi_{\mu\nu}(k+(p_2-p_1))-\Pi_{\mu\nu}(k)\right\}
 \nonumber \\
  &\simeq&  \frac{\partial}{\partial p_4}
  \Pi_{\mu\nu}(k+p)
  \simeq  P^T_{\mu\nu} \frac{\partial}{\partial p_4}        
  \Pi_M (k+p).
\eea
This relation is also a direct consequence of the Ward identity 
for the HDL gluon three-point function. Using equ.~(\ref{gam_p4})
and continuing to euclidean space we get the following expression 
for the non-abelian vertex function in the limit $(l_{p_1}-l_{p_2})
\ll(p_1-p_2)_4$ 
\be
\Gamma^{(nab)}_\alpha(p) = g^3C^{(nab)}v_\alpha\int\frac{dl_4}{2\pi}
 \int\frac{l^2dl}{4\pi^2} \int dx\, 
  \frac{1-x^2}{[l_4^2+l^2+\frac{\pi}{2}m^2\frac{l_4}{l}]^2}
  \frac{\pi}{2}m^2\frac{1}{l}  \frac{1}{i(l_4+p_4)-lx}.
\ee
This integral can be evaluated using the same strategy 
as for the abelian vertex. First we observe that to leading
logarithmic accuracy we can replace $(1-x^2)\to 1$. The 
integral over $x$ then gives
\be 
\int dx\, \frac{1}{i(l_4+p_4)-lx} = 
 \frac{2i}{l}\arctan \left( \frac{l}{l_4+p_4}\right).
\ee
In the limit $p_4\to 0$ the integral over $l_4$ is dominated 
by the discontinuity of the inverse tan function and
\be 
\label{vertex_7} 
\lim_{(p_1)_4\to (p_2)_4} \lim_{l_{p_1}\to l_{p_2}}
 \Gamma^{(nab)}_\alpha (p_1,p_2)  = 
\frac{g^3C^{(nab)} v_\alpha}{12\pi^2}
\log\left(\frac{\Lambda}{p_4}\right).
\ee
The non-abelian vertex function in the limit $(l_{p_1}-l_{p_2})
\gg(p_1-p_2)_4$ can be computed using the same methods. We find 
that there is no logarithmic enhancement near the Fermi surface. 
We conclude that the abelian and non-abelian soft dense loop 
vertex functions have the same logarithmic singularity and that 
the sum of the two contributions is proportional to $C^{(ab)}+
C^{(nab)}=C_F$. 

\section{Color Superconductivity}
\label{sec_gap}

 Soft dense loop corrections to the fermion self energy 
and the quark-gluon vertex function become comparable to
the free propagator and the free vertex at the scale $E\sim 
\mu \exp(-1/g^2)$. This implies that at this scale soft 
dense loops have to be resummed. Physically, this 
resummation corresponds to the study of non-Fermi liquid
effects in dense quark matter \cite{Boyanovsky:2000bc}.
However, before non-Fermi liquid effects become important
the quark-quark interaction in the BCS channel becomes 
singular. The scale of superfluidity is $E\sim \mu\exp(-1/g)$
\cite{Son:1999uk}. This scale is smaller than the HDL scale 
but larger than the SDL scale. This implies that we have 
to resum quark-quark scattering with HDL dressed gluon 
propagators but that SDL corrections to the quark self 
energy and the quark-gluon vertex are small and can be 
treated perturbatively. 

 The resummation of the quark-quark scattering amplitude
in the BCS channel leads to the formation of a non-zero 
gap in the single particle spectrum. We can take this
effect into account in the high density effective theory 
by including a tree level gap term 
\be 
\label{gap}
{\cal L} = \Delta\, R_l^\Gamma(\vec{v}\cdot \hat\Delta)
  \psi_{-v}\sigma_2\Gamma \psi_{v} + h.c..
\ee
Here, $\Gamma$ is any of the helicity structures introduced
in Sect.~\ref{sec_hdet}, $R_l^\Gamma(x)$ is the corresponding 
angular factor and $\hat\Delta$ is a unit vector. The magnitude 
of the gap is determined variationally, by requiring the free 
energy to be stationary order by order in perturbation theory. 
We shall see that the gap varies with the energy or the residual 
momentum on a scale set by the gap itself. As a result the energy 
dependence is relevant and we cannot replace the gap by its value 
on the Fermi surface. In order to do perturbation theory in 
the presence of a gap term we will use the Nambu-Gorkov 
method and introduce a two component field $\Psi=(\psi_v,
\psi^\dagger_{-v}\sigma_2)$. The inverse propagator for the 
field $\Psi$ is 
\be 
\label{ng}
S^{-1}(p) = \left( \begin{array}{cc}
v\cdot p & \Delta(p) R_l^\Gamma(\vec{v}\cdot\hat\Delta)\Gamma \\ 
\Delta^*(p)  R_l^\Gamma(\vec{v}\cdot\hat\Delta)\Gamma^\dagger & 
\bar{v}\cdot p \end{array}\right) ,
\ee
where $\bar{v}_\mu = (1,-\vec{v})$. The variational principle
for the gap $\Delta$ gives the Dyson-Schwinger equation 
\be 
\label{gap_1}
\Delta(p_4) =\frac{2g^2}{3} \int \frac{d^4q}{(2\pi)^4} 
 \frac{\Delta(q_4)}{q_4^2+l_q^2+\Delta(q_4)^2}
  v_\mu v_\nu D_{\mu\nu}(p-q) ,
\ee
where the factor in front of the integral is the color factor
corresponding to the color anti-symmetric $[\bar{3}]$ channel. 
It is sufficient to solve this equation to leading logarithmic 
accuracy. We shall see in the next section that corrections to 
the gap can be computed perturbatively, without solving the gap 
equation to higher accuracy. To leading logarithmic accuracy
the gap equation is dominated by the IR divergence in the 
magnetic gluon propagator. This IR divergence is independent 
of the helicity and angular momentum channel. We have 
\be 
\label{gap_2}
\Delta(p_4) =\frac{g^2}{18\pi^2}
\int_0^{\Lambda_{\|}} dq_4 
\frac{\Delta(q_4)}{\sqrt{q_4^2+\Delta(q_4)^2}}
\log\left( \frac{\Lambda_\perp}{|p_4^2-q_4^2|^{1/2}}\right).
\ee
The leading logarithmic behavior is independent of the ratio 
of the cutoffs and we can set $\Lambda_{\|}=\Lambda_\perp
=\Lambda$. We introduce the dimensionless variables 
variables $x=\log(2\Lambda/(q_4+\epsilon_q))$ and $y
=\log(2\Lambda/(p_4+\epsilon_p)$ where $\epsilon_q=(q_4^2
+\Delta(q_4))^{1/2}$. In terms of dimensionless variables 
the gap equation is given by
\be 
\label{gap_3}
\Delta(y) = \frac{g^2}{18\pi^2}\int_0^{x_0} dx\, \Delta(x) K(x,y),
\ee
where $x_0=\log(2\Lambda/\Delta_0)$ and $K(x,y)$ is the kernel 
of the integral equation. At leading order we can use the 
approximation $K(x,y)=\min(x,y)$ \cite{Son:1999uk}. We can 
perform an additional rescaling $x=x_0\bar{x}$, $y=x_0\bar{y}$. 
Since the leading order kernel is homogeneous in $x,y$ we can 
write the gap equation as an eigenvalue equation 
\be 
\label{gap_4}
\Delta(\bar{y}) = x_0^2 \frac{g^2}{18\pi^2}\int_0^1 d\bar{x}\,
\Delta(\bar{x}) K(\bar{x},\bar{y}),
\ee
where the gap function is subject to the boundary conditions
$\Delta(0)=0$ and $\Delta'(1)=0$. Son observed that equ.~(\ref{gap_4})
is equivalent to the differential equation \cite{Son:1999uk}
\be 
\label{gap_5}
\Delta''(\bar{x}) = -x_0^2 \left(\frac{g^2}{18\pi^2}\right) 
  \Delta(\bar{x}) ,
\ee
which has the solutions 
\be 
\label{gap_6}
\Delta_n(\bar{x}) = \Delta_{n,0} 
\sin\left( \frac{g}{3\sqrt{2}\pi}x_{0,n}\bar{x} \right), 
\hspace{1cm}
x_{0,n}= (2n+1)\frac{3\pi^2}{\sqrt{2}g}.
\ee
The physical solution corresponds to $n=0$ which gives the 
largest gap, $\Delta_0=2\Lambda \exp(-3\pi^2/(\sqrt{2}g))$.
Solutions with $n\neq 0$ have smaller gaps and are not 
global minima of the free energy. 

\section{Perturbative corrections to the gap parameter}
\label{sec_pgap}

 The complete set of solutions of the leading order gap equation 
can be used to set up a perturbative scheme for computing corrections 
to the gap function. This perturbative scheme is similar to the one 
used by Brown et al.~in order to compute corrections to the critical 
temperature \cite{Brown:1999yd}. We first observe that the 
functions $\Delta_n(\bar{x})$ form an orthogonal set of solutions
of the gap equation
\be 
\label{ortho}
\int_0^1 d\bar{x}\,\Delta_n(\bar{x})\Delta_m(\bar{x}) = 
 \delta_{m,n}.
\ee
We now consider the gap equation with the kernel $K(x,y)+\delta K(x,y)$
where $K(x,y)$ contains the leading IR divergence and $\delta K(x,y)$ is
a perturbative correction. We have 
\be 
\Delta(\bar{y}) = x_0^2 \frac{g^2}{18\pi^2}\int_0^1 d\bar{x}\,
\Delta(\bar{x}) K(\bar{x},\bar{y}) 
 +  x_0 \frac{g^2}{18\pi^2}\int_0^1 d\bar{x}\,
\Delta(\bar{x}) \delta K(x_0\bar{x},x_0\bar{y}).
\ee
We write the gap function $\Delta(\bar{x})$ and the eigenvalue 
$x_0$ as a perturbative expansion in $\delta K$ 
\bea
\Delta(\bar{x}) &=& \Delta^{(0)}(\bar{x}) + \Delta^{(1)}(\bar{x}) 
 +  \Delta^{(2)}(\bar{x}) + \ldots , \\
 x_0 &=& x_0^{(0)} + x_0^{(1)} + x_0^{(2)} + \ldots .
\eea
Using the orthogonality of the unperturbed solutions 
$\Delta_n(\bar{x})$ we can derive expressions for $\Delta^{(i)}$
and $x_0^{(i)}$. These expressions are very similar to 
ordinary Rayleigh-Schroedinger perturbation theory. At first 
order in perturbation theory we get 
\be 
\label{eval_1}
x_0^{(1)} = -\frac{1}{2}\left( x_0^{(0)}\right)^2
 \frac{g^2}{18\pi^2}
\int_0^1d\bar{x}\int_0^1d\bar{y}\, \Delta_0^{(0)}(\bar{x}) 
  \delta K(x_0\bar{x},x_0\bar{y}) \Delta_0^{(0)}(\bar{y}),
\ee
and 
\be 
\label{gfct_1}
c^{(1)}_k = 
 \frac{1}{1-\left(\frac{1}{2k+1}\right)^2}\,
 x_0^{(0)}  \frac{g^2}{18\pi^2}
\int_0^1d\bar{x}\int_0^1 d\bar{y}\,
 \Delta^{(0)}_0(\bar{x}) \delta K(x_0\bar{x},x_0\bar{y}) 
 \Delta^{(0)}_k(\bar{y}), 
\ee
where $c_k^{(1)}$ are the components of $\Delta^{(1)}(\bar{x})$
in the basis of the unperturbed gap functions, $\Delta^{(1)}(\bar{x}) 
= \sum c^{(1)}_k \Delta^{(0)}_k(\bar{x})$. Equ.~(\ref{eval_1})
shows that the first order correction to the pairing gap can 
be computed using the unperturbed gap function. 

The simplest example for a correction to the kernel is 
is a contact term 
\be 
\label{pert_ct}
\delta K(x_0\bar{x},x_0\bar{y}) = \log(b).
\ee
In Sect.~\ref{sec_match} we will show that the coefficient 
$b$ is determined by four-fermion operators in the effective theory. 
The unperturbed gap is given by $\Delta=\Lambda \exp(-x^{(0)}_0)=
\Lambda\exp(-3\pi^2/(\sqrt{2}g))$. Using equ.~(\ref{eval_1})
and equ.~(\ref{pert_ct}) we get 
\be 
\label{gap_ct}
  \Delta= \Lambda e^{-(x^{(0)}_0+x^{(1)}_0+\ldots)}
 = b\Lambda e^{-\frac{3\pi^2}{\sqrt{2}g}}.
\ee
We conclude that contact terms modify the prefactor of the 
gap. We also notice that because of the orthogonality
of the gap functions a contact term does not modify 
the shape of the gap function. 

 We can also study the effect of the quark self energy
\cite{Brown:1999aq,Wang:2001aq}. Equ.~(\ref{sigma_3})
implies a wave function renormalization
\be
Z(p_4) = 1-\frac{g^2C_F}{12\pi^2}
  \log\left( \frac{\Lambda}{p_4} \right) .
\ee
The corresponding correction to the kernel of the gap equation
is 
\be 
\label{pert_zf}
\delta K(x_0\bar{x},x_0\bar{y}) = 
 -\frac{g^2}{9\pi^2} \left(x_0\bar{x}\right)
  K(x_0\bar{x},x_0\bar{y}),
\ee
where we have used $C_F=4/3$ for $N_c=3$ and $K(x,y)=\min(x,y)$ 
is the leading order kernel. Using equ.~(\ref{eval_1}) we find
\bea 
x^{(1)}_0 &=& \frac{1}{2} 
 \left( x_0^{(0)} \right)^2  \frac{g^2}{18\pi^2}
 \int_0^1d\bar{x}\int_0^1 d\bar{y}\, 
\frac{g^2}{9\pi^2}  (x_0\bar{x})
 \Delta^{(0)}_0(\bar{x}) K(x_0\bar{x},x_0\bar{y})
  \Delta^{(0)}_0(\bar{y}) \nonumber \\
\label{gap_zf}
 & =& \frac{1}{2}\frac{g^2}{9\pi^2} 
 \left( x_0^{(0)} \right)^2
 \int_0^1d\bar{x}\, \bar{x} \left[\Delta^{(0)}_0(\bar{x})\right]^2
 = \frac{4+\pi^2}{8},
\eea
where we have used the fact that $\Delta^{(0)}_0(\bar{x})$ is 
an eigenfunction of the unperturbed kernel. The result 
equ.~(\ref{gap_zf}) corresponds to a reduction of the 
gap by a factor $\exp(-(4+\pi^2)/8)$. We can also see
that the fermion self energy correction leads to an 
admixture of higher harmonics of the gap function. 
However, these admixtures are small, $c_k^{(1)}=O(g)$,
for all values of $k$.

 There is a slight subtlety with regard to the result 
equ.~(\ref{gap_zf}). In section \ref{sec_sloop} we 
computed the wave function renormalization in the normal 
phase. We observed, however, that the scale of non-Fermi 
liquid effects, $E\sim \mu\exp(-1/g^2)$, is exponentially 
small as compared to the scale where pairing sets in, 
$E\sim \mu\exp(-1/g)$. This implies that also the normal
component of the fermion self energy should be computed 
with the gap taken into account. It is easy to see that 
in this case equ.~(\ref{sigma_3}) is modified to 
\be
\label{sigma_4}
\Sigma(p_4)  \simeq  \frac{g^2C_F}{12\pi^2}p_4 \log\left( 
  \frac{\Lambda}{\sqrt{p_4^2+\Delta^2}}\right) .
\ee 
Pairing removes the infrared divergence in the 
wave function renormalization. However, in the superfluid
phase the fermion self energy is only modified for very 
small energies $p_4<\Delta$ whereas the correction to the 
gap is dominated by the regime $p_4>\Delta$. As a consequence 
the result equ.~(\ref{gap_zf}) is not changed. Finally, we have 
to consider the role of soft dense loop vertex corrections 
given in equ.~(\ref{vertex_5}) and (\ref{vertex_7}). The 
infrared logarithm in the vertex correction only appears in 
the regime $|l_p-l_q|\ll |p_4-q_4|$. Since the gap equation 
determines the anomalous self energy on the quasi-particle
mass shell, $p_4\simeq (l_q^2+\Delta^2)^{1/2}$, this 
implies that $l_q\ll q_4$. However, this condition 
eliminates the BCS logarithm in the gap equation (\ref{gap_1}).
As a consequence, the infrared logarithm in the quark-gluon
vertex does not modify the eigenvalue $x_0$ at $O(1)$ in 
the coupling constant. Vertex corrections to the gap 
equation were first considered in \cite{Schafer:1999jg}
but the arguments given in that work are not correct. 
A more detailed study can be found in \cite{Brown:2000eh}.

\section{Matching contact terms}
\label{sec_match}

 Consider the effect of the $O(1/\mu^2)$ contact term 
equ.~(\ref{v_bcs_0}) on the pairing gap. The contribution of 
this term to the kernel of the gap equation involves a sum over 
patches, see Fig.~\ref{fig_sc2}. As a consequence the correction 
to the kernel, $g^2\delta K(x,y) = 6 V_0^{++}$, is not suppressed 
by $1/\mu^2$. In the previous section we showed that $O(g^2)$ 
constants in the kernel contribute to the eigenvalue at $O(1)$. 
This implies that we have to determine to coefficients of 
$O(g^2/\mu^2)$ contact terms in the high density effective 
theory in order to determine the eigenvalue of the gap equation 
to $O(1)$. 

 This can be achieved by matching the quark-quark 
scattering amplitude in the BCS channel. The tree level
scattering amplitude in the spin zero color anti-triplet
channel is given by
\be 
\label{f_tree}
 f(\theta) = \frac{2g^2}{3} \left\{
 \frac{\frac{1}{2}(1+\cos\theta)}
   {2\mu^2(1-\cos\theta)+\Pi_E} + 
 \frac{\frac{1}{2}(3-\cos\theta)}
   {2\mu^2(1-\cos\theta)+\Pi_M} \right\}.
\ee
At leading order in the effective theory this amplitude 
is represented by 
\bea
\label{f_hdet}
  f_{HDET}(\theta) &=& \frac{2g^2}{3} \left.\left\{
 \frac{1}
   {2\mu^2(1-\cos\theta)+\Pi_E} + 
 \frac{1}
   {2\mu^2(1-\cos\theta)+\Pi_M} \right\}\right|_{1-\cos\theta
   <\Lambda_\perp^2/(2\mu^2)} \nonumber \\
 & & \hspace{1cm}\mbox{}
 + \frac{V^{++}_l(\Lambda_\perp)}{\mu^2}P_l(\cos\theta),
\eea
where the collinear term is cut off at $l_\perp^2=\Lambda_\perp^2$.
The matching condition requires that the partial wave amplitudes
corresponding to equ.~(\ref{f_tree}) and (\ref{f_hdet}) are the same. 
Since the collinear term in the high density effective theory is 
regulated by a UV cutoff the counterterm depends on the cutoff, 
too. The matching condition is simplest for $\Lambda_\perp^2
=2\mu^2$. The s-wave term is given by
\be
V^{++}_0(2\mu^2) = \frac{2g^2}{3}\frac{1}{4} \int_{-1}^1 dx\,
 \left\{ \frac{-\frac{1}{2}(1-x)}{(1-x)+\Pi_E/(2\mu^2)}
 + \frac{\frac{1}{2}(1-x)}{(1-x)+\Pi_M/(2\mu^2)} \right\}
 = 0,
\ee
up to corrections of $O(g^4)$. The cutoff dependence of 
$V_0^{++}$ is controlled by the renormalization group equation 
\be 
\label{v0_rge}
\Lambda_\perp^2\frac{d}{d\Lambda_\perp^2}V_0^{++}(\Lambda_\perp^2)
 = \frac{g^2}{3}.
\ee
This equation implies that the kernel in the gap equation is 
independent of the cutoff $\Lambda_\perp$. We find  
\cite{Schafer:1999jg,Hong:2000fh,Pisarski:2000tv}
\be 
\label{k_ct}
 K(p_4,q_4) = \log\left( \frac{\Lambda_\perp}{|p_4^2-q_4^2|^{1/2}}\right)
 + \log\left(\frac{512\pi^4\mu}{g^5\Lambda_\perp}
   \left(\frac{2}{N_f}\right)^{\frac{5}{2}}\right).
\ee
Using the leading order result equ.~(\ref{gap_6}), the 
perturbative corrections equ.~(\ref{gap_ct},\ref{gap_zf}), and
the value of the counterterm given in equ.~(\ref{k_ct}) 
we get the standard result for the gap in the 2SC phase
\be 
\label{gap_final}
\Delta = 512\pi^4(2/N_f)^{5/2}\mu g^{-5}
 e^{-\frac{4+\pi^2}{8}}e^{-\frac{3\pi^2}{\sqrt{2}g}}.
\ee
Note that at this order we did not have to fix the dependence 
of the counterterms on the longitudinal cutoff $\Lambda_{\|}$. 

 We saw that $V_0^{++}(2\mu^2)$ vanishes to leading order 
in perturbation theory. This is not the case for higher 
partial wave terms and operators with non-zero spin. For
example, the total angular momentum one counter term
in the total helicity zero channel is
\be 
V^{++}_1(2\mu^2) = \frac{2g^2}{3}\frac{3}{4} \int_{-1}^1 dx\,
\frac{x[(\frac{1}{2}+\frac{x}{2}) 
   +    (\frac{3}{2}-\frac{x}{2})]-2}{1-x} 
 = -6\frac{g^2}{3} .
\ee
The tree level matrix element in the helicity one channel 
is proportional to $(1+\cos\theta)/2$ for both electric 
and magnetic gluon exchanges. The corresponding counterterm 
is 
\be
V^{+-}_1(2\mu^2) = \frac{2g^2}{3}\frac{3}{4} \int_{-1}^1 dx\,
\frac{d^{(1)}_{11}(x)[(\frac{1}{2}+\frac{x}{2}) 
                    + (\frac{1}{2}+\frac{x}{2})]-2}{1-x} 
 = -\frac{9}{2}\frac{2g^2}{3} ,
\ee
where we have used $d^{(1)}_{11}(x)=(1+x)/2$. These results 
imply that the total angular momentum one gaps are given 
by $\Delta^{++}_1=\exp(-6)\Delta^{++}_0$ and $\Delta^{+-}_1
= \exp(-9/2)\Delta^{++}_0$ where $\Delta^{++}_0$ is the 
s-wave gap given in equ.~(\ref{gap_final}) 
\cite{Brown:1999yd,Schafer:2000tw,Schmitt:2002sc}.

\section{Summary}
\label{sec_sum}

  In this work we studied several problems related to the use 
of effective field theories in QCD at large baryon density. 
We showed that the power counting in $1/\mu$ is complicated 
by the appearance of two scales inside loop integrals. Hard 
dense loops involve the large scale $\mu^2$ and lead to 
phenomena such as screening and damping at the scale $g\mu$. 
Soft loops only involve small scales and lead to superfluidity 
and non-Fermi liquid behavior at exponentially small
scales. We also showed that contact terms in the effective 
lagrangian are suppressed by powers of $1/\mu^2$, but 
they get enhanced by hard loops. As a result contact 
terms contribute to the pre-exponent of the pairing 
gap at $O(1)$. We performed the necessary matching 
calculation to determine four-fermion operators in the 
effective theory.

  There are many problems that remain to be studied. We
would like to understand not only how to count powers of 
$E/\mu$ but also logarithms. This is essential for 
understanding the structure of the expansion of the 
gap and other quantities in the coupling constant. Another 
issue is the choice of regularization scheme. In this 
work we have used a momentum space cutoff throughout. There 
are, however, some HDET calculations that have been performed 
using dimensional regularization. It is clearly important 
to understand the power counting in both schemes. 

 We would also like to have a better understanding of 
gauge invariance. The high density effective theory 
has the advantage that the pre-exponent of the gap 
is determined by counterterms that are matched against
on-shell scattering amplitudes and that are therefore
manifestly gauge invariant. This result agrees with 
explicit calculations in a generalized Coulomb gauge
\cite{Pisarski:2001af} but not with calculations in 
a general covariant gauge \cite{Hong:2000fh}. A formal
argument that the full quasi-particle propagator is gauge 
invariant was recently presented in \cite{Gerhold:2003js}.
It is well known that the hard dense loop effective action 
equ.~(\ref{S_hdl}) is gauge invariant. It is not clear 
whether there is a similar statement for soft dense loops. 
Brown et al.~showed that the soft quark self energy and 
quark gluon vertex function satisfy BRST identities
\cite{Brown:2000eh}. We have also checked that the 
IR divergences in the soft wave function renormalization 
and vertex are independent of the gauge parameter in 
a general Coulomb gauge. On the other hand, Hsu et 
al.~have argued that one can find a non-local gauge 
in which there is no wave function renormalization
\cite{Hong:2003ts}.

 Hong and Hsu have argued that the high density effective 
theory can be formulated non-perturbatively, using a lattice 
regulator, and that the leading order theory has a positive 
euclidean measure \cite{Hong:2002nn}. Such a formulation 
might be useful in order to study certain non-perturbative 
questions and to provide guidance in the search for 
algorithms capable of simulating gauge theories at non-zero
baryon density. Our results suggest, however, that there are
some difficulties with this proposal. The leading order 
effective theory is defined on a single patch, or, if a gap 
term is added, on two patches corresponding to Fermi velocities 
$\pm\vec{v}$. In this theory the ground state is likely a coherent 
particle-hole state of the type suggested by Deryagin, Grigoriev 
and Rubakov \cite{Deryagin:rw}. The reason is that the forward 
scattering amplitude in the particle-hole channel is larger than 
the one in the particle-particle channel. This state is disfavored
in the full theory, because the corresponding four-fermion
operator is not enhanced by hard dense loops. This means that 
operators that are superficially sub-leading are important 
in selecting the correct ground state. Physically, this is 
related to the fact that the particle-hole state can only 
become coherent over the entire Fermi surface if the surface
is nested, but not if the Fermi surface is spherical. 
Another problem is how to correctly
incorporate the boundary conditions for the effective theory. 
The lagrangian of the leading order theory is equal to the 
non-relativistic QCD (NRQCD) lagrangian. However, the physics
of the high density theory and NRQCD are very different. For
example, there is no screening in the gluon two-point function
in NRQCD. This issue is also related to the necessity to 
add a two-gluon counterterm to the effective theory. In 
perturbation theory this difference is encoded in different 
$i\epsilon$ prescriptions for the fermion propagator. It would 
be interesting to understand how the boundary conditions are 
realized on a euclidean lattice. Some of these issues are 
studied in \cite{Hong:2003zq}.
 
Acknowledgments: We would like to thank S.~Beane, P.~Bedaque,
and D.~Rischke for useful discussions. This work was supported 
in part by US DOE OJI grant and by US DOE grant DE-FG-88ER40388. 


\newpage

\begin{figure}
\leavevmode
\begin{center}
\includegraphics[width=15cm,angle=0]{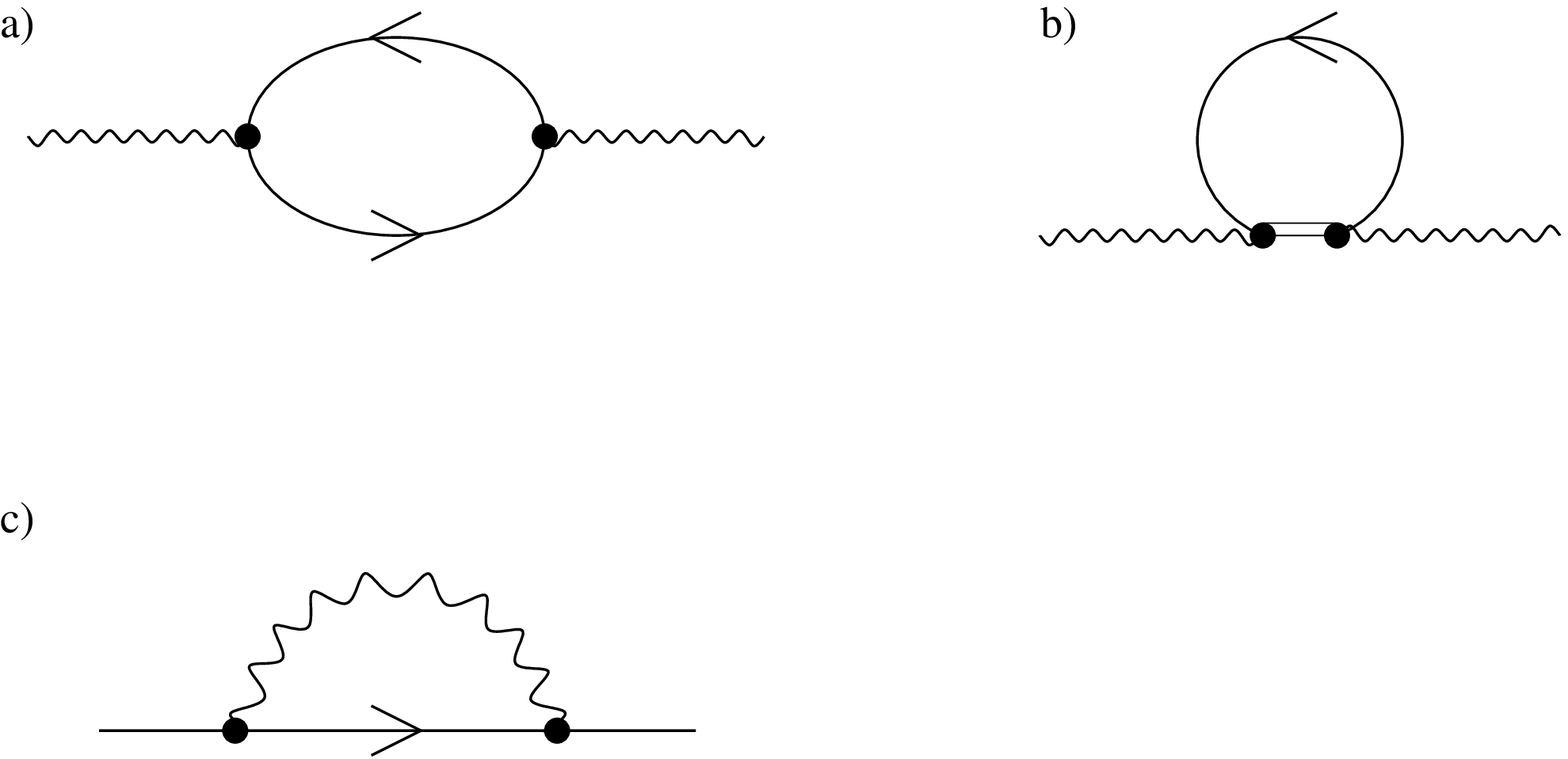}
\end{center}
\caption{\label{fig_hdl1}
Hard dense loop contributions to the the gluon 
and quark self energy. Fig a) shows the particle-hole
contribution to the gluon self energy, b) shows the
anti-particle contribution, and c) shows the quark
self energy.}
\end{figure}

\begin{figure}
\leavevmode
\begin{center}
\includegraphics[width=15cm,angle=0]{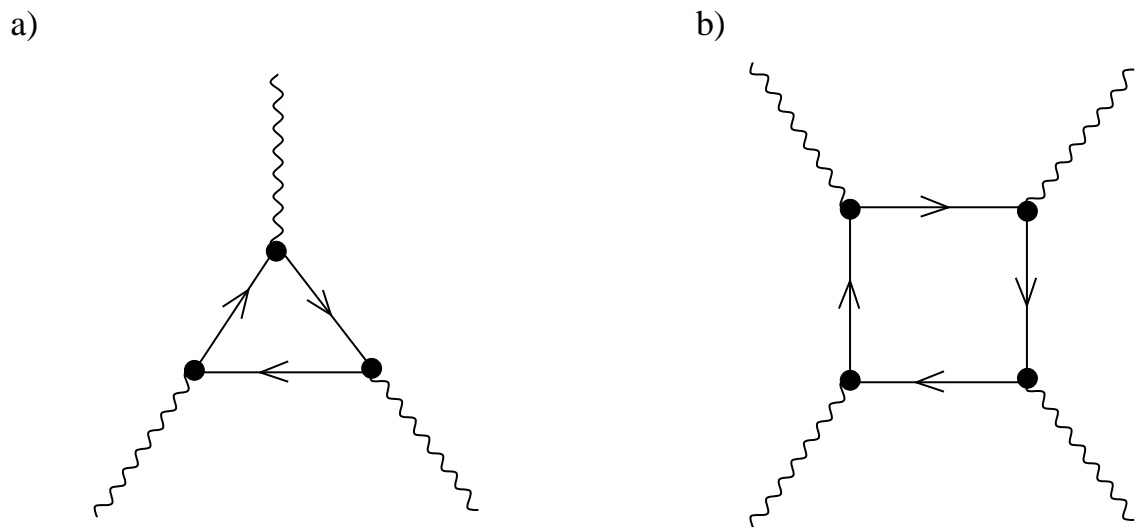}
\end{center}
\caption{\label{fig_hdl3}
 Hard dense loop contributions to the gluon three 
and four function.}
\end{figure}

\begin{figure}
\leavevmode
\begin{center}
\includegraphics[width=15cm,angle=0]{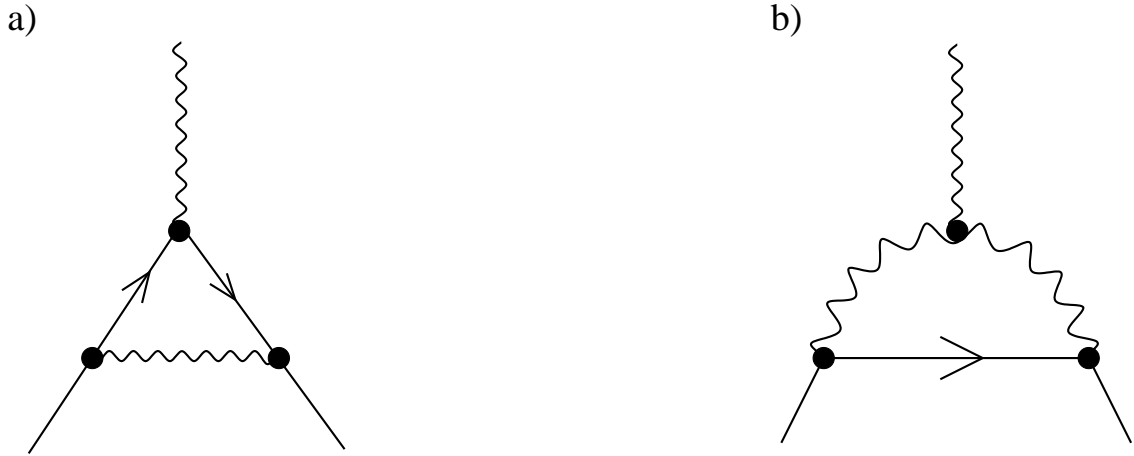}
\end{center}
\caption{\label{fig_hdl2}
 Hard dense loop contribution to the quark-gluon vertex 
function. Figure a) shows the abelian vertex and figure
b) the non-abelian vertex.}
\end{figure}

\begin{figure}
\leavevmode
\begin{center}
\includegraphics[width=12cm,angle=0]{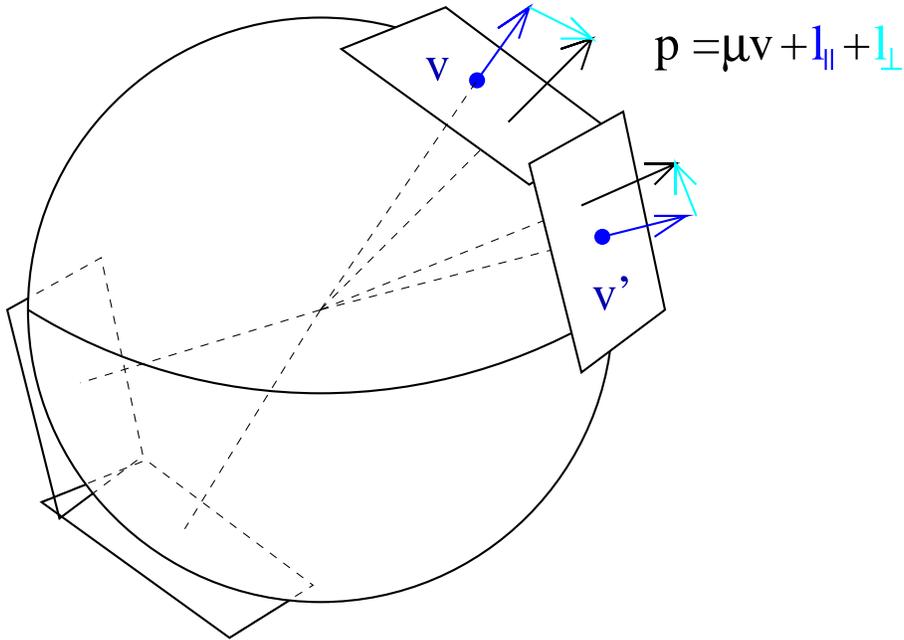}
\end{center}
\caption{\label{fig_fs}
High density effective field theory description of 
excitations near the Fermi surface. The effective 
theory is defined on patches labeled by the local
Fermi velocity $v$. Momenta are decomposed with 
respect to $v$, $\vec{p}=\mu\vec{v}+l_\perp+l_{||}$. }
\end{figure}

\begin{figure}
\leavevmode
\begin{center}
\includegraphics[width=9cm,angle=0]{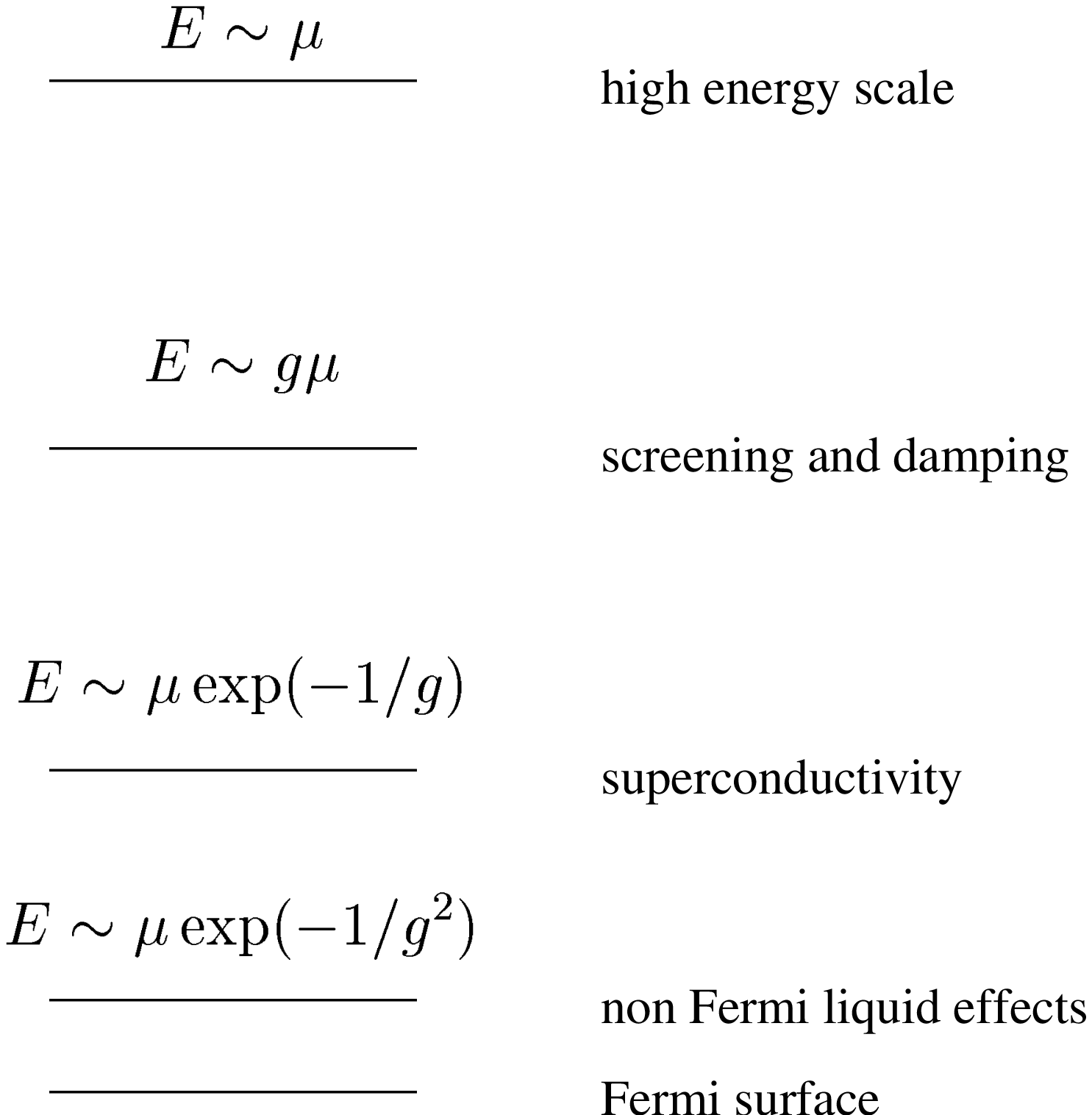}
\end{center}
\caption{\label{fig_scales}
Scales of the high density effective field theory. }
\end{figure}

\begin{figure}
\leavevmode
\begin{center}
\includegraphics[width=6cm,angle=0]{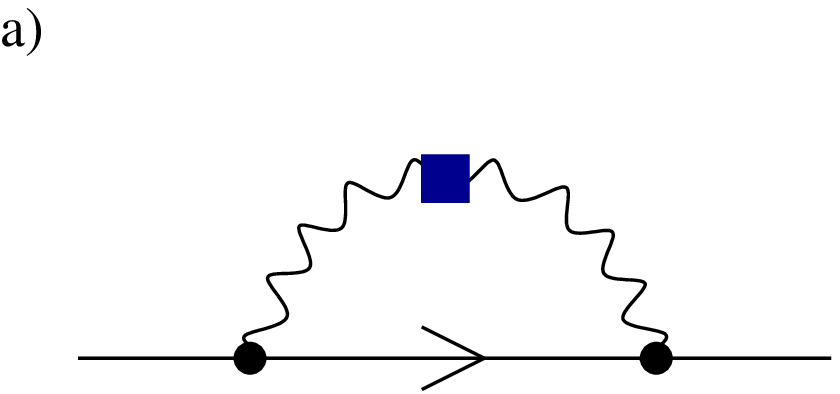}

\vspace*{1cm}
\includegraphics[width=12cm,angle=0]{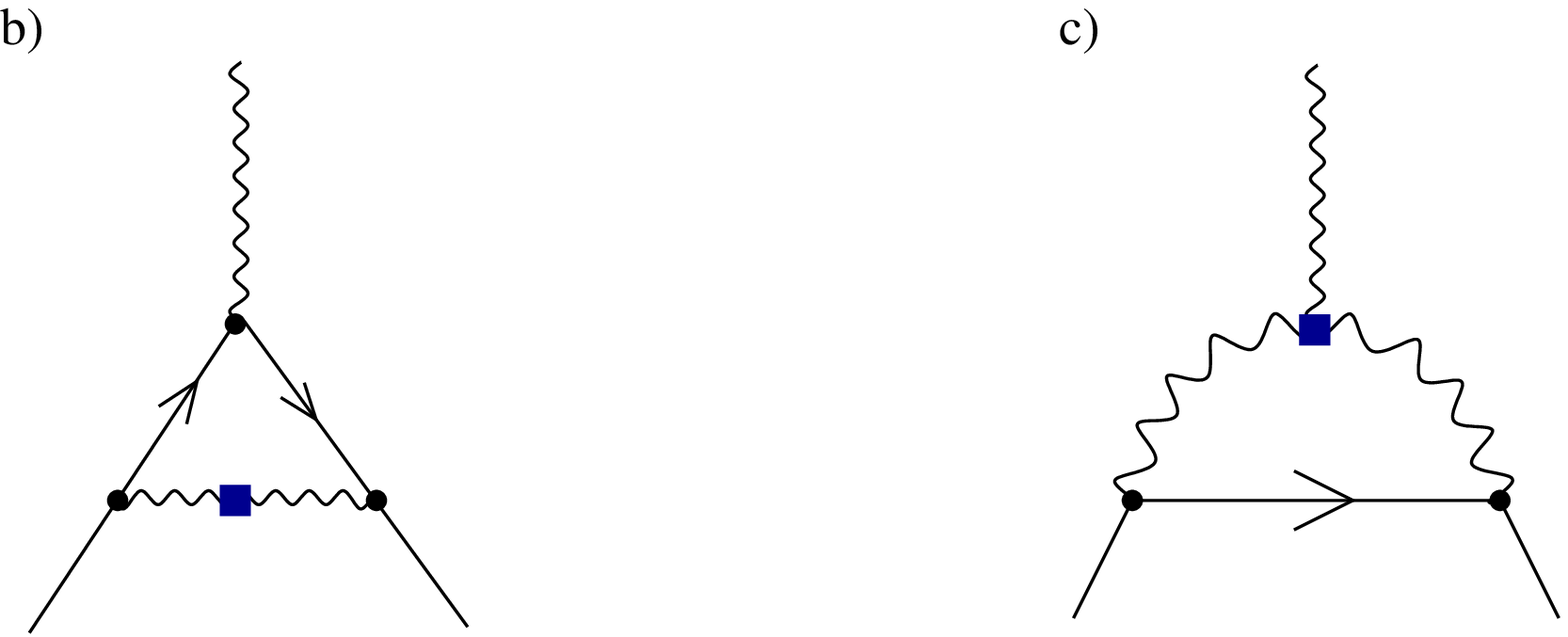}
\end{center}
\caption{\label{fig_sdl}
Soft dense loop contribution to the quark self 
energy (Fig.~a)) and the quark-gluon vertex 
function (Figs.~b) and c)). The solid squares
denote the HDL gluon self energy and three-point
function.}
\end{figure}

\begin{figure}
\leavevmode
\begin{center}
\includegraphics[width=6cm,angle=0]{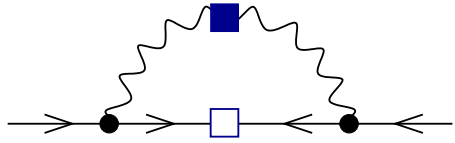}
\end{center}
\caption{\label{fig_sc1}
Leading order gap equation for the superconducting gap
in the high density effective theory. The solid square 
denotes the HDL self energy and the open square is the 
anomalous quark self energy. }
\end{figure}

\begin{figure}
\leavevmode
\begin{center}
\includegraphics[width=12cm,angle=0]{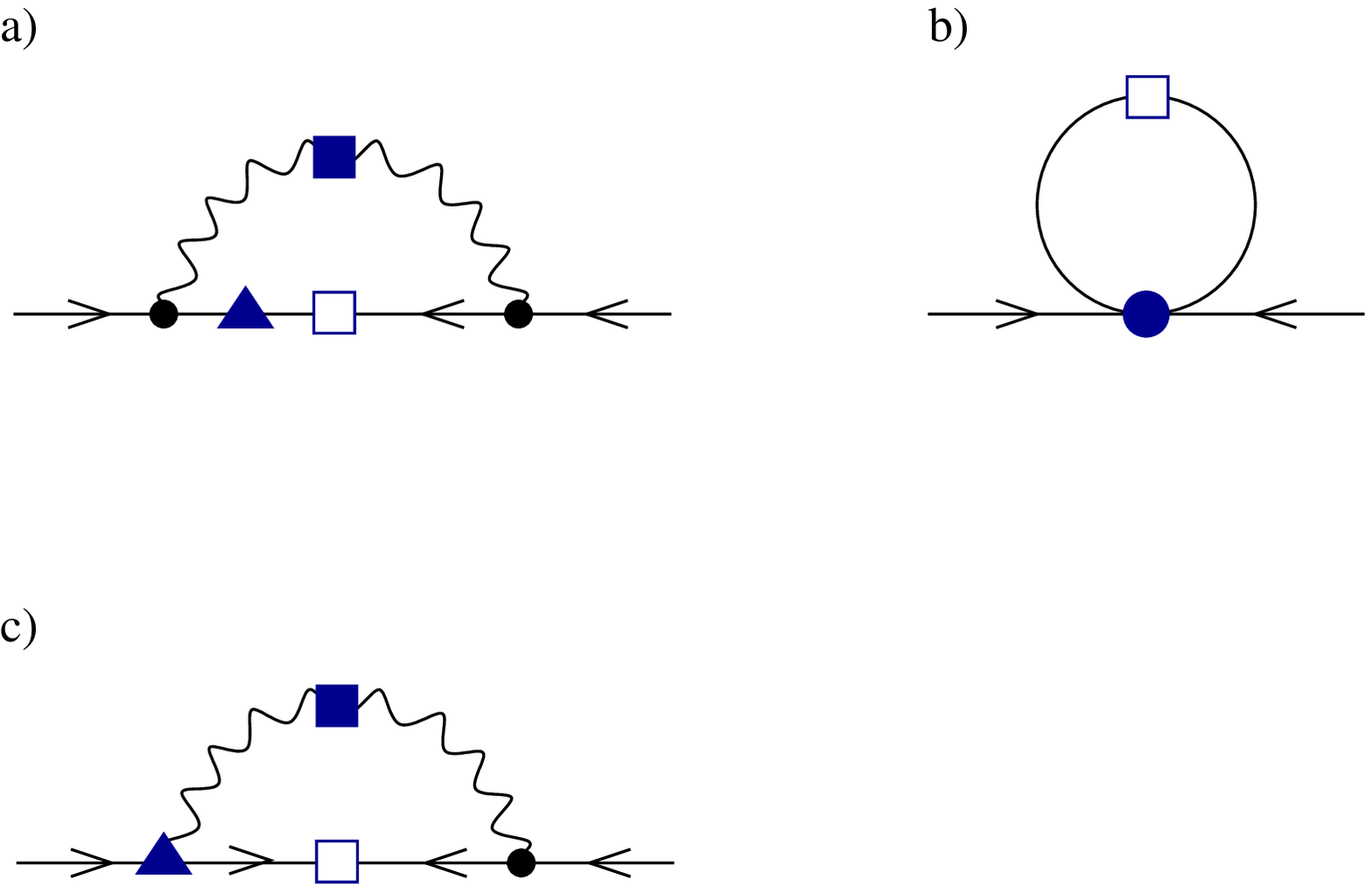}
\end{center}
\caption{\label{fig_sc2}
Corrections to the gap equation. Fig.~a) shows the 
soft dense loop quark self energy correction, Fig.~b) 
local four-fermion interactions, and Fig.~c) shows 
the soft dense loop vertex correction. The solid triangles
denote the SDL self energy and vertex correction and the 
solid circle is the four-fermion interaction in the effective
theory.}
\end{figure}

\begin{thebibliography}{20}

\bibitem{Bailin:1984bm}
D.~Bailin and A.~Love,
Phys.\ Rept.\  {\bf 107}, 325 (1984).
 
\bibitem{Alford:1998zt}
M.~Alford, K.~Rajagopal and F.~Wilczek,
Phys.\ Lett.\  {\bf B422}, 247 (1998)
[hep-ph/9711395].
 
\bibitem{Rapp:1998zu}
R.~Rapp, T.~Sch{\"a}fer, E.~V.~Shuryak and M.~Velkovsky,
Phys.\ Rev.\ Lett.\  {\bf 81}, 53 (1998)
[hep-ph/9711396].

\bibitem{Alford:1999mk}
M.~Alford, K.~Rajagopal and F.~Wilczek,
Nucl.\ Phys.\  {\bf B537}, 443 (1999)
[hep-ph/9804403].

\bibitem{Rajagopal:2000wf}
K.~Rajagopal and F.~Wilczek,
The condensed matter physics of QCD, in:
Festschrift in honor of B.L. Ioffe, 'At the Frontier 
of Particle Physics / Handbook of QCD', M. Shifman, ed., 
World Scientific, Singapore, 
hep-ph/0011333.

\bibitem{Alford:2001dt}
M.~Alford, 
Ann.\ Rev.\ Nucl.\ Part.\ Sci.\  {\bf 51}, 131 (2001)
[hep-ph/0102047].

\bibitem{Schafer:2003vz}
T.~Sch{\"a}fer,
Quark Matter, BARC workshop on Quarks and Mesons; 
to appear in the proceedings,
hep-ph/0304281.

\bibitem{Rischke:2003mt}
D.~H.~Rischke,
The Quark-Gluon Plasma in Equilibrium, 
to appear in Prog. Part. Nucl. Phys, 
nucl-th/0305030.

\bibitem{Reddy:2002ri}
S.~Reddy,
Acta Phys.\ Polon.\ B {\bf 33}, 4101 (2002)
[nucl-th/0211045].

\bibitem{Alford:wf}
M.~Alford,
Lect.\ Notes Phys.\  {\bf 583}, 81 (2002).

\bibitem{Schafer:1999ef}
T.~Sch{\"a}fer and F.~Wilczek,
Phys.\ Rev.\ Lett.\  {\bf 82}, 3956 (1999)
[hep-ph/9811473].

\bibitem{GellMann:1957}
M.~Gell-Mann, K.~Brueckner, 
Phys.\ Rev.\ {\bf 106}, 364 (1957).

\bibitem{Holstein:1973}
T.~Holstein, A.~E.~Norton, P.~Pincus,
Phys.\ Rev.\ {\bf B8}, 2649 (1973).

\bibitem{Reizer:1989}
M.~Yu.~Reizer, 
Phys.\ Rev.\ {\bf B 40}, 11571 (1989).

\bibitem{Hong:2000tn}
D.~K.~Hong,
Phys.\ Lett.\ B {\bf 473}, 118 (2000)
[hep-ph/9812510].

\bibitem{Hong:2000ru}
D.~K.~Hong,
Nucl.\ Phys.\ B {\bf 582}, 451 (2000)
[hep-ph/9905523].

\bibitem{Beane:2000ji}
S.~R.~Beane and P.~F.~Bedaque,
Phys.\ Rev.\ D {\bf 62}, 117502 (2000)
[nucl-th/0005052].

\bibitem{Beane:2000ms}
S.~R.~Beane, P.~F.~Bedaque and M.~J.~Savage,
Phys.\ Lett.\ B {\bf 483}, 131 (2000)
[hep-ph/0002209].

\bibitem{Schafer:2001za}
T.~Sch{\"a}fer,
Phys.\ Rev.\ D {\bf 65}, 074006 (2002)
[hep-ph/0109052].

\bibitem{Casalbuoni:2001ha}
R.~Casalbuoni, R.~Gatto, M.~Mannarelli and G.~Nardulli,
Phys.\ Lett.\ B {\bf 524}, 144 (2002)
[hep-ph/0107024].

\bibitem{Nardulli:2002ma}
G.~Nardulli,
Riv.\ Nuovo Cim.\  {\bf 25N3}, 1 (2002)
[hep-ph/0202037].

\bibitem{Braaten:1989mz}
E.~Braaten and R.~D.~Pisarski,
Nucl.\ Phys.\ B {\bf 337}, 569 (1990).

\bibitem{Blaizot:1993bb}
J.~P.~Blaizot and J.~Y.~Ollitrault,
Phys.\ Rev.\ D {\bf 48}, 1390 (1993)
[hep-th/9303070].

\bibitem{Manuel:1995td}
C.~Manuel,
Phys.\ Rev.\ D {\bf 53}, 5866 (1996)
[hep-ph/9512365].

\bibitem{Rischke:2000qz}
D.~H.~Rischke,
Phys.\ Rev.\ D {\bf 62}, 034007 (2000)
[nucl-th/0001040].

\bibitem{Schafer:1999na}
T.~Sch{\"a}fer and F.~Wilczek,
Phys.\ Lett.\  {\bf B450}, 325 (1999)
[hep-ph/9810509].

\bibitem{Schafer:2002ty}
T.~Sch\"afer,
Phys.\ Rev.\ D {\bf 65}, 094033 (2002)
[hep-ph/0201189].

\bibitem{Jacob:at}
M.~Jacob and G.~C.~Wick,
Annals Phys.\  {\bf 7}, 404 (1959)
[Annals Phys.\  {\bf 281}, 774 (2000)].

\bibitem{Braaten:1991gm}
E.~Braaten and R.~D.~Pisarski,
Phys.\ Rev.\ D {\bf 45}, 1827 (1992).

\bibitem{Braaten:1992jj}
E.~Braaten,
Can.\ J.\ Phys.\  {\bf 71}, 215 (1993)
[hep-ph/9303261].

\bibitem{Brown:1999yd}
W.~E.~Brown, J.~T.~Liu and H.~c.~Ren,
Phys.\ Rev.\ D {\bf 62}, 054016 (2000)
[hep-ph/9912409].

\bibitem{Brown:2000eh}
W.~E.~Brown, J.~T.~Liu and H.~c.~Ren,
Phys.\ Rev.\ D {\bf 62}, 054013 (2000)
[hep-ph/0003199].

\bibitem{Boyanovsky:2000bc}
D.~Boyanovsky and H.~J.~de Vega,
Phys.\ Rev.\ D {\bf 63}, 034016 (2001)
[hep-ph/0009172].

\bibitem{Manuel:2000nh}
C.~Manuel,
Phys.\ Rev.\ D {\bf 62}, 114008 (2000)
[hep-ph/0006106].

\bibitem{Vanderheyden:1996bw}
B.~Vanderheyden and J.~Y.~Ollitrault,
Phys.\ Rev.\ D {\bf 56}, 5108 (1997)
[hep-ph/9611415].

\bibitem{Son:1999uk}
D.~T.~Son,
Phys.\ Rev.\ D {\bf 59}, 094019 (1999)
[hep-ph/9812287].

\bibitem{Brown:1999aq}
W.~E.~Brown, J.~T.~Liu and H.~Ren,
Phys.\ Rev. {\bf D61}, 114012 (2000)
[hep-ph/9908248].

\bibitem{Wang:2001aq}
Q.~Wang and D.~H.~Rischke,
Phys.\ Rev.\ D {\bf 65}, 054005 (2002)
[nucl-th/0110016].

\bibitem{Schafer:1999jg}
T.~Sch{\"a}fer and F.~Wilczek,
Phys.\ Rev.\  {\bf D60}, 114033 (1999)
[hep-ph/9906512].
  
\bibitem{Hong:2000fh}
D.~K.~Hong, V.~A.~Miransky, I.~A.~Shovkovy and L.~C.~Wijewardhana,
Phys.\ Rev.\  {\bf D61}, 056001 (2000)
[hep-ph/9906478].

\bibitem{Pisarski:2000tv}
R.~D.~Pisarski and D.~H.~Rischke,
Phys.\ Rev.\  {\bf D61}, 074017 (2000)
[nucl-th/9910056].

\bibitem{Schafer:2000tw}
T.~Sch{\"a}fer,
Phys.\ Rev.\ D {\bf 62}, 094007 (2000)
[hep-ph/0006034].

\bibitem{Schmitt:2002sc}
A.~Schmitt, Q.~Wang and D.~H.~Rischke,
Phys.\ Rev.\ D {\bf 66}, 114010 (2002)
[nucl-th/0209050].

\bibitem{Pisarski:2001af}
R.~D.~Pisarski and D.~H.~Rischke,
Nucl.\ Phys.\ A {\bf 702}, 177 (2002)
[nucl-th/0111070].

\bibitem{Gerhold:2003js}
A.~Gerhold and A.~Rebhan,
preprint, 
hep-ph/0305108.

\bibitem{Hong:2003ts}
D.~K.~Hong, T.~Lee, D.~P.~Min, D.~Seo and C.~Song,
preprint, hep-ph/0303181.

\bibitem{Hong:2002nn}
D.~K.~Hong and S.~D.~Hsu,
Phys.\ Rev.\ D {\bf 66}, 071501 (2002)
[hep-ph/0202236].

\bibitem{Deryagin:rw}
D.~V.~Deryagin, D.~Y.~Grigoriev and V.~A.~Rubakov,
Int.\ J.\ Mod.\ Phys.\ A {\bf 7}, 659 (1992).

\bibitem{Hong:2003zq}
D.~K.~Hong and S.~D.~Hsu,
preprint, hep-ph/0304156.



\end{thebibliography}
\end{document}